\documentclass[prb,superscriptaddress,twocolumn]{revtex4-1}
\usepackage{times,amsmath}
\usepackage{epsfig}
\usepackage{color}
\usepackage{longtable}
\usepackage{graphicx}
\usepackage{dcolumn}
\usepackage{bm}
\usepackage{tabularx}
\usepackage{hyperref}
\usepackage{multirow}
\hypersetup{colorlinks=true, citecolor=blue, filecolor=blue, linkcolor=blue, urlcolor=blue}

\begin{document}
\title{Computational Design of Flexible Electrides with Non-trivial Band Topology}

\author{Sheng-Cai Zhu}
\affiliation{Department of Physics and Astronomy, High Pressure Science and Engineering Center, University of Nevada, Las Vegas, NV 89154, USA}

\author{Lei Wang}
\affiliation{Shenyang National Laboratory for Materials Science, Institute of Metal Research, Chinese Academy of Sciences, School of Materials Science and Engineering, University of Science and Technology of China, 110016, Shenyang, China}

\author{Jing-Yu Qu}
\affiliation{Department of Physics and Astronomy, High Pressure Science and Engineering Center, University of Nevada, Las Vegas, NV 89154, USA}
\affiliation{College of Science, China Agricultural University, Beijing, 100083, China}

\author{Jun-Jie Wang}
\affiliation{Materials Research Center for Element Strategy, Tokyo Inst. of Technology, 4259-SE3 Nagatsuta-cho, Midori-ku, Yokohama, Kanagawa, 226-8501, Japan.}

\author{Timofey Frolov}
\affiliation{Lawrence Livermore National Laboratory, Livermore, CA, 94550, USA}

\author{Xing-Qiu Chen}
\email{xingqiu.chen@imr.ac.cn}
\affiliation{Shenyang National Laboratory for Materials Science, Institute of Metal Research, Chinese Academy of Sciences, School of Materials Science and Engineering, University of Science and Technology of China, 110016, Shenyang, China}

\author{Qiang Zhu}
\email{qiang.zhu@unlv.edu}
\affiliation{Department of Physics and Astronomy, High Pressure Science and Engineering Center, University of Nevada, Las Vegas, NV 89154, USA}

\date{\today}
\begin{abstract}
Electrides, with their excess electrons distributed in crystal cavities playing the role of anions, exhibit a variety of unique electronic and magnetic properties. In this work, we employ the first-principles crystal structure prediction to identify a new prototype of A$_3$B electride in which both interlayer spacings and intralayer vacancies provide channels to accommodate the excess electrons in the crystal. This A$_3$B type of structure is calculated to be thermodynamically stable for two alkaline metals oxides (Rb$_3$O and K$_3$O). Remarkably, the unique feature of multiple types of cavities makes the spatial arrangement of anionic electrons highly flexible via elastic strain engineering and chemical substitution, in contrast to the previously reported electrides characterized by a single topology of interstitial electrons. More importantly, our first-principles calculations reveal that Rb$_3$O is a topological Dirac nodal line semimetal, which is induced by the Rb-$s$ $\rightarrow$ O-$p$ band inversion at the general electronic k momentums in the Brillouin zone associated with the intersitial electric charges. The discovery of flexible electride in combining with topological electronic properties opens an avenue for electride design and shows great promises in electronic device applications.
\end{abstract}
\maketitle

\section{Introduction}
Electrides are important functional materials with exotic electronic properties and many potential applications, such as in optoelectrics \cite{Hosono-PNAS-2017} and catalysis \cite{Kitano-NChem-2012, Inoue-Catalysis-2016, Kim-CC-2015}. The key feature of electride lies in its non-bound electrons serving as anions in the crystal \cite{Dye-Science-1990, Allan-EPL-1990}. Such electron distribution picture is significantly different from the conventional covalent and ionic compounds \cite{Dye-ACR-2009}: 
(i) it violates the standard valence rule for the excess unsaturated electrons trapped around the interstitial sites, while the conventional solids have zero formal charges. 
(ii) it localizes the anionic electrons into the interstitial sites as non-nucleus-bound electrons, while in conventional compounds electrons are distributed around atoms to form bonds; 
(iii) it has stoichiometric composition, unlike the random defects of color centers. Electrides with the high mobility of the non-nucleus-bound anionic electrons are candidates of superconductor \cite{Miyakawa-JACS-2007, Hosono-PC-2009, Hosono-PTRSA-2015}, while the “excess” electrons are excellent catalysts for ammonia and alkynes synthesis due to their high electron donor ability \cite{Kim-CC-2015, Lu-JACS-2016, Kitano-NC-2015, Kobayashi-CCT-2017, Kanbara-JACS-2015}. Recently, it was also suggested that the so-called unique floating electrons in the electrides are favorable for achieving band inversions needed for topological phase transition in the electronic states \cite{Hirayama-PRX-2018}.

The search for room temperature stable electride materials has been a long pursuit in materials research since Dye's pioneering work \cite{Ellaboudy-1983-JACS, Dye-ACR-2009}. The first crystalline organic electride was made by Dye and coworkers at 1983 \cite{Ellaboudy-1983-JACS}. Several others have been synthesized by Dye's group \cite{Redko-JACS-2005, Dye-ACR-2009}. However, the application of organic electrides was limited by their thermal instability. To improve the stability, the electride studies have been shifted to inorganic materials recently. The first room-temperature stable inorganic electrided C12A7:2e$^-$ was synthesized by Matsuishi et al \cite{Matsuishi-Science-2003} through selectively removing two oxygen ions per unit cell from the parent compound mayenite (Ca$_6$Al$_7$O$_{16}$) via oxygen-reducing processes. In the cubic unit cell of C12A7:2e-, four excess electrons are trapped to twelve crystallographically equivalent zero-dimensional (0D) cages. Due to its low work function \cite{Toda-AM-2007} and high electron mobility \cite{Matsuishi-Science-2003}, C12A7:2e$^-$ can efficiently promote the catalytic activity for ammonia synthesis \cite{Kitano-NChem-2012, Kuganathan-JACS-2014, Hayashi-JACS-2014}, and it shows promise as a low electron-injection barrier material for organic light-emitting diodes (OLED) fabrication as well \cite{Hosono-PNAS-2017}. 

The mobility of the interstitial electrons strongly depends on the topology of the cavities confining the anionic electrons. In a 0D electride, the anionic electrons are localized in lattice cavities and isolated from each other, while in 1D, 2D and even 3D electrides the electrons are connected along lattice channels or planes. Electrides with higher degree of connectivity are more desirable for catalytic and other applications. Ca$_2$N, being predicted to possess intrinsic nucleus-free 2D electrons gas at the interlayer space \cite{Steinbrenner-JPCS-1998}, has been prepared by Gregory and coworkers \cite{gregory2000dicalcium}, and later confirmed to be a 2D electride by Lee and coworkers \cite{Lee-Nature-2013}. With the anionic electrons directly exposed to the environment, Ca$_2$N is able to greatly promote the ammonia synthesis \cite{Inoue-Catalysis-2016,Kobayashi-CCT-2017}. Recently, several 1D-like electrides (La$_8$Sr$_2$(SiO$_4$)$_6$ \cite{Zhang-JPCL-2015}, Nb$_5$Ir$_3$ \cite{Zhang-QM-2017}, Y$_5$Si$_3$ \cite{Lu-JACS-2016} and Sr$_5$P$_3$ \cite{Wang-JACS-2017}) have been established both in theory and experiment. It was also proposed that materials with the same chemical composition may exhibit different electron confinements from 0D, 2D to 3D due to structural diversity \cite{Tsuji-JACS-2016}. Moreover, the recent high pressure studies substantially expanded our understanding on electrides. Under compression, many simple metals (Li \cite{Pickard-PRL-2009, Matsuoka-Nature-2009}, Na \cite{Ma-Nature-2009}, K \cite{Pickard-PRL-2011}, and compounds (NaHe$_2$\cite{Dong-NChem-2017} and Mg$_3$O$_2$ \cite{Zhu-PCCP-2013}) were suggested to adopt structures with valence electrons localized in the interstitial regions. Nevertheless, these high pressure electrides are not recoverable under ambient conditions. 

To date, a library of new electride materials, with interstitial electrons forming different channel topologies, have been proposed. From the viewpoint of device application, it is desirable to have a material whose properties could be altered by external conditions or fields in a controllable manner. A recent work demonstrated that the spin-alignment of interstitial electrons in 2D Y$_2$C can be effectively tuned by introducing the isovalent Sc substitution on the Y site \cite{Park-JACS-2017-tune}. It was also reported that the magnetic state of organic Cs$^+$(15-crown-5)$_2$e$^-$ electride could be changed by external pressure \cite{dale2018pressure}. However, it is still unknown whether the properties could be modulated to a greater extent. Our knowledge about controlling the non-bound-electrons topology is surprisingly poor, which hinders the rational design of electrides with desired properties. 
    
A key to realize the modulation is to find a material in which the distribution of interstitial electrons has flexible responses to external fields. This motivates us to perform a thorough crystal structure search to find such materials. In this work, we combined first-principles evolutionary crystal structure prediction \cite{Oganov-JCP-2006, Lyakhov-CPC-2013} with high throughput screening methods (from online database of Materials Project \cite{MP-2013}) based on density functional theory (DFT), which allow us to effectively explore the structure spaces with unknown structural prototypes and calculate their stability maps in a variety of chemical systems. For the first time, we identified the monoclinic ($C$2/$m$) form of A$_3$B with anti-MoCl$_3$ prototype as a thermodynamically stable (or marginally stable) electride in metal oxides (Rb$_3$O and K$_3$O) at ambient pressure condition, as well as a group of layered structures analogical to transition metal trichlorides (MCl$_3$) with similar thermodynamic stability. The distribution of the interstitial electrons in $C$2/$m$-Rb$_3$O can be readily modulated by applying strain or chemical pressure. Our further calculations reveal that Rb$_3$O is a topological Dirac nodal-line semimetal. We believe the discovery of flexible electride with combined toplogical properties will show great promises in future electronic devices applications.

\section{COMPUTATIONAL METHODS}
\subsection{Crystal Structure Prediction} 
We started our investigation with alkaline metal oxides, which are likely candidates for metal rich compounds to hold anionic interstitial electrons. In particular, we were interested in heavy alkaline metals (Rb, Cs), because these metals exhibit rich oxidation chemistry and form a number of stable metal-rich compounds according the online database of Materials Project (examples of such compounds include Rb$_6$O, Rb$_9$O$_2$, Cs$_7$O, Cs$_{11}$O$_3$ and Cs$_3$O). To screen for other possible compounds not listed in the database, we performed a first-principles crystal structure prediction (CSP) on Rb-O and Cs-O systems using the USPEX code \cite{Oganov-JCP-2006, Lyakhov-CPC-2013}. In this search, all combinations of Rb and O in the unit cell were allowed (within the limitation that the total number of atoms does not exceed 32). The first generation of structures was created with random symmetries and random compositions. All structures were relaxed at ambient pressure and zero temperature, at the level of density functional theory (DFT) as implemented in the VASP code \cite{Vasp-PRB-1996}. In the type of crystal structure search with variable compositions, it is not the total energy, but the largest normalized enthalpy (or free energy at zero temperature) of formation from other compounds in the system, that defines the fitness function. This construction leads to the so-called convex hull diagram. According to the fitness ranking, the worst structures (40\%) were discarded and new generations were created by heredity, mutation and random generator operations. The best structure from each generation was kept. All structure optimizations evolved over maximum of 30 generations. 

\subsection{High-throughput Screening}
While the first-principles CSP can perform a thorough search of all possible structures and different compositions, it is computationally expensive and can be only conducted in a limited set of chemical systems. To improve the efficiency of materials discovery, we take advantage of the existing high-throughput data mining approaches based on available materials databases, which already contain information about crystal structures and energies calculated for a variety of compounds, making it unnecessary to repeat the expensive DFT calculations. Combining our CSP approach with resources from Materials Project database \cite{MP-2013} can accelerate the discovery of new electrides. We selected the best ten structures by fitness ranking from both CSP runs on both Rb-O and Cs-O, and then applied chemical substitution to other binaries in group IA-VIA and IIA-VA. Together with the available data entries in Materials Project \cite{MP-2013}, we computed the final convex hull diagram for each binary system under investigation.

\subsection{DFT Calculations} 
All DFT calculations were performed using the plane wave code VASP \cite{Vasp-PRB-1996} based on the projector augmented wave scheme. The exchange-correlation functional was described by the generalized gradient approximation in the Perdew-Burke-Ernzerhof parameterization (GGA-PBE) \cite{PBE-PRL-1996}, and the energy cutoff of the plane wave was set as 1000 eV. We used Rb PAW potential with a [Ar] core of 2.541 a.u. and 9 valence electrons (4s$^2$4p$^6$5s$^1$), O PAW potential with a [He] core of 1.550 a.u and 6 valence electrons (2s$^2$2p$^4$). In order to account for the van der Waals (vdW) interaction, all energies were further recomputed using the optPBE functional \cite{Klimevs-PRB-2011}. The geometry convergence criterion was set as 0.001 eV/{\AA} for the maximal component of force and 0.001 GPa for stress. The electronic structures are calculated based on DFT. The Brillouin zone (BZ) was sampled by uniform $\Gamma$-centered meshes with the reciprocal space resolution of 2$\pi$ $\times$ 0.03 {\AA$^{-1}$}, which was verified to be accurate enough for the systems under investigation. The dynamic stability of the structures was carefully checked by the phonon spectra calculation with the finite displacement method as implemented in Phonopy code \cite{Togo-PRB-2008}. For the electronic and phonon band structure calculations, the band paths are taken based on the suggestion from Ref. \cite{HINUMA2017140}. A tight binding model Hamiltonian based on the maximally localized Wanner functions (MLWFs) \cite{Marzari-PRB-1997, Souza-PRB-2001} has been constructed in the basis of $s$ and $p$ orbits of Rb atom and $p$ orbits of O atom. In the Wanner representations, we applied an iterative Green functions method \cite{Lopez-JPF-1984, Lopez-JPF-1985} based on the bulk tight-binding model to obtain the surface density of states and the surface projected Fermi surface of the semi-infinite system neglecting possible surface reconstructions and charge rearrangements.

\subsection{Electride Analysis}
We identified electrides by calculating electron localization function (ELF) \cite{ELF}. If the ELF of a given material yields non-nuclear maxima and the electrons can be assigned to ionic bonding, we consider the materials as a possible electride. Moreover, we performed a detailed band structure and partial charge analysis to identify the topology of the interstitial electrons. Since we are interested in only energetically favorable electreides, the ELF analysis was performed on a subset of the generated structures, which were stable or marginally stable with respect to the constructed convex hull. 

\section{Results}
In the following sections, we will first discuss the results of Rb-O system obtained from first-principles CSP method. In particular, we introduce a new monoclinic ($C2/m$) form of Rb$_3$O with anti-MoCl$_3$ prototype as a thermodynamically stable electride at ambient pressure condition from our prediction and demonstrate that it is an electride which is found to possess flexible anionic electrons under elastic strain and chemical substitution. Then, we discuss its intriguing topological properties. Last, we extend it to other binary systems via high throughput data mining method and discuss the statistics of binary inorganic electrides.

\begin{figure}
\epsfig{file=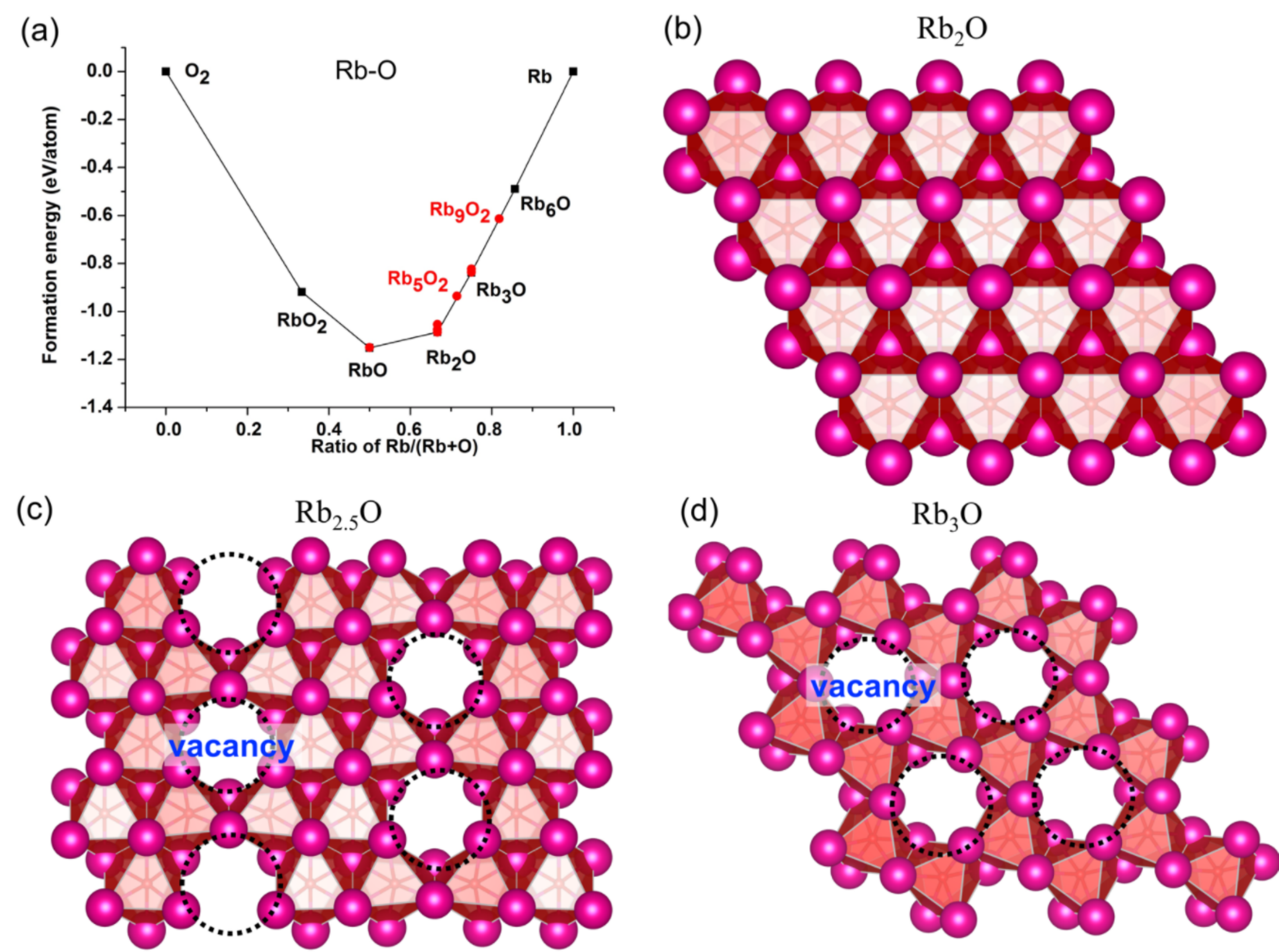, width=0.45\textwidth}
\caption{\label{PhaseDiagram} {(a) Convex hull diagram of the Rb-O system at opt-PBE level. Red circles denote metastable compounds; black squares represent stable compounds. The top view of monolayers in layered structures of $R$-3$m$ Rb$_2$O (b), $C$2/$m$ Rb$_5$O$_2$ (c) and $C$2/$m$ Rb$_3$O (d), respectively.}}
\end{figure}

\subsection{Phase Diagram of Rb-O system}
Through sampling more than thousands of structures at ambient pressure from evolutionary CSP on Rb-O, we obtained a series of stable and metastable structures. Fig. \ref{PhaseDiagram}a shows the formation enthalpies of predicted structures relative to the elemental Rb and O$_2$. At PBE level, RbO$_2$, RbO, Rb$_2$O, Rb$_5$O$_2$, Rb$_3$O, Rb$_9$O$_2$ and Rb$_6$O, were calculated to be stable, see Fig. S1 \cite{SI}. Among them, Rb$_3$O and Rb$_5$O$_2$ were newly predicted in this work, while the rest of them have been already reported in the database. This example illustrated the power of CSP method in identifying new compounds with unknown stoichiometries. Including vdW interaction within optPBE functional does not qualitatively change the relative stability. Instead, it only slightly shifted Rb$_5$O$_2$ and Rb$_9$O$_2$ from being stable to being marginally stable (8 meV/atom and 3 meV/atom above the hull, respectively).

Fig. \ref{PhaseDiagram}b-d illustrates structures of Rb$_2$O, Rb$_5$O$_2$ and Rb$_3$O. Large magenta atoms correspond to Rb, while smaller red atoms represent oxygen. The Rb-O bonds are also shown in red. All three compounds have layered structures and they differ only in the number of oxygen vacancies. Rb$_2$O has an anti-CaCl$_2$ type structure with $R$-3$m$ symmetry (the experimental Rb$_2$O is in anti-fluorite type, which was calculated to be 2 meV/atom less stable than $R$-3$m$ structure at PBE level, but 10 meV/atom more stable at optPBE level). In this structure, every oxygen is octahedrally coordinated with Rb atoms. Each (001) basal plane is composed of Rb$_6$O octahedral connected along their edges. Similar to Rb$_2$O, Rb$_5$O$_2$ also has a layered structure with the $C$2/$m$ monoclinic symmetry. To account for the different composition, 'O-vacancies' are present in the (001) basal planes. Thus, the Rb$_{10}$O$_4$ layer can be obtained by removing one oxygen out of every five units of Rb$_2$O. Similarly, the layered structure of $C$2/$m$ Rb$_3$O can be constructed from $R$-3$m$ Rb$_2$O, by removing one oxygen out of every three units of Rb$_2$O. This transformation also lowers the symmetry of the structure from trigonal to monoclinic. 

According to the formal charges of Rb (+1) and O (-2), all the compounds between Rb$_2$O and Rb$_3$O should have excess electrons. $R$-3$m$ Rb$_2$O has a balanced charge which means it is unlikely to be an electride. On the other hand, Ca$_2$N, with the same structural type, has +1 charge per formula unit and becomes electride with the excess electrons confined to the interlayer space. Given the structural similarity, Rb$_5$O$_2$ and Rb$_3$O are likely to be electrides. Moreover, both Rb$_5$O$_2$ and Rb$_3$O have O-vacancies in the basal plane, which might provide another type of sites to accommodate the electrons. In our crystal structural search, we also observed several other low-energy layered Rb$_3$O structures similar to $C$2/$m$ Rb$_3$O with different inter-layer stacking. The detailed structure information is shown in Fig. S3 \cite{SI} and their properties are summarized in Table S1. Due to the strong structural similarity, they are energetically competitive and may share similar physical properties with $C$2/$m$ Rb$_3$O. Since $C$2/$m$ Rb$_3$O is stable under both PBE and optPBE functionals, we shall focus on this compound in the proceeding sections.

\begin{figure*}
\epsfig{file=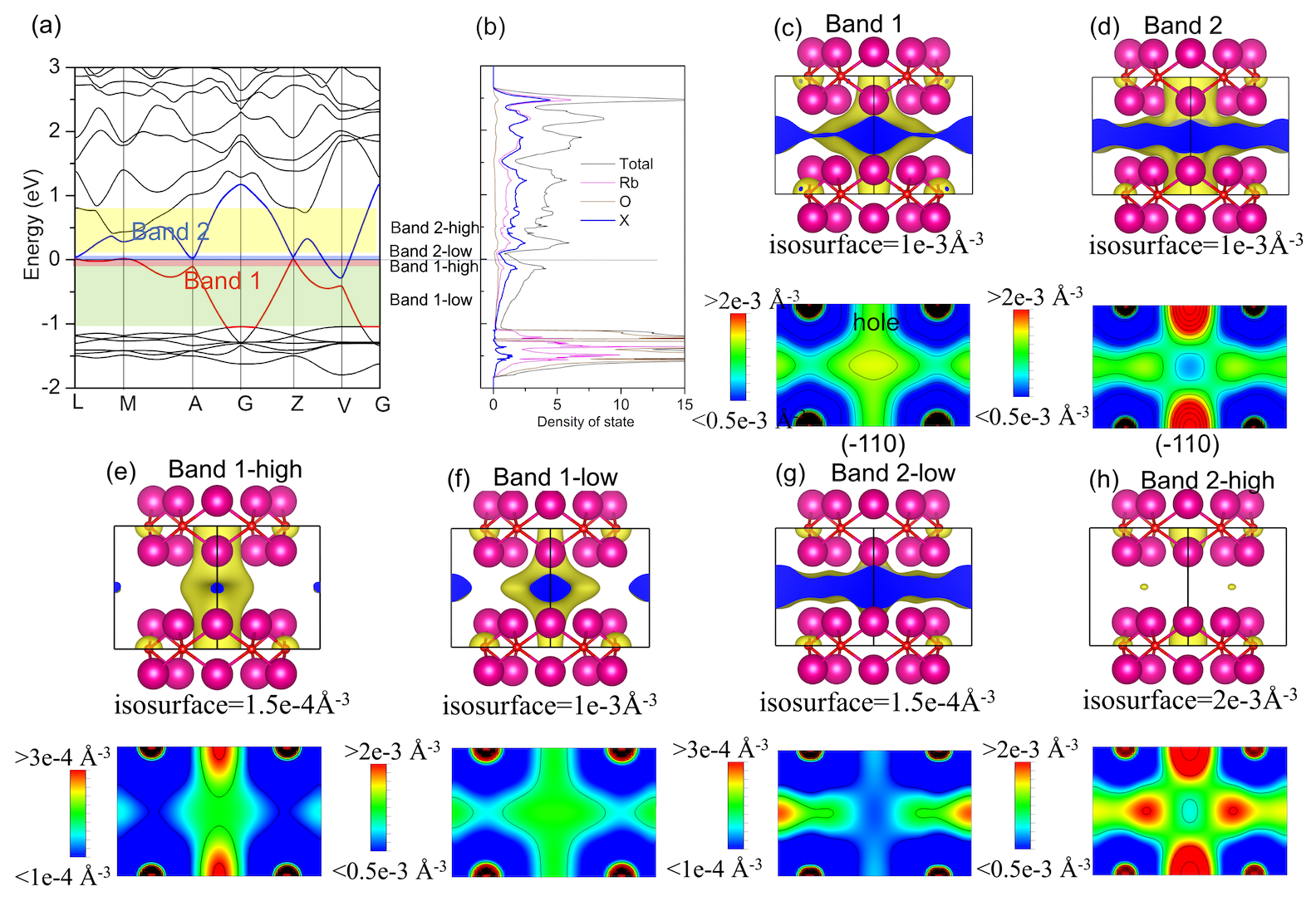, width=0.9\textwidth}
\caption{\label{band} {The electronic band structure (a), partial density of states (b), the isosurface and map of partial charge density distribution of interstitial Band 1 (c),  Band 2 (d), the isosurface and map of density distribution of interstitial Band 1 (e-f) and Band 2 (g-h) of $C$2/$m$ Rb$_3$O phase. The interstitial bands, mainly contributed by the interstitial electrons, are highlighted in bold red and blue lines. In c-h, the large magenta spheres denote Rb atoms, and the small red spheres denote O atoms.}}
\end{figure*}

\subsection{Electronic Structure of Rb$_3$O}
We performed a detailed analysis of the electronic properties of $C$2/$m$ Rb$_3$O. Specifically, we calculated its band structure, density of states (DOS), the decomposed partial charge distribution and ELF (see Fig. S5 \cite{SI}). As shown in Fig. \ref{band}a-d, we found that the layered $C$2/$m$ Rb$_3$O structure is a typical electride. There are two bands crossing the Fermi level ($E_{\textrm{Fermi}}$). As a result, its DOS forms two peaks spanning around $E_{\textrm{Fermi}}$. In order to identify the contributions from interstitial electrons, we placed pseudoatoms with a Wigner-Seitz radius of 2.00 {\AA} at interlayer interstitial site and intralayer vacancies, and then projected the portions of the wave functions within these spheres to obtain the partial density of state (PDOS) curves for the interstitial sites. Fig. \ref{band}b shows that the DOS accumulation near the  $E_{\textrm{Fermi}}$ is mainly from the non-atom-centered orbitals located around the interstitial site while the contributions from the atomic orbitals of Rb and O are much smaller.

The existence of the interstitial electrons is further confirmed by the decomposed partial charge densities according to the energy (Fig. \ref{band}c-d). By plotting the partial charge density of the two bands which cross $E_{\textrm{Fermi}}$, namely Band 1 and Band 2, we can clearly find that they correspond to interstitial electrons. The picture of two excess electrons forming two partially occupied interstitial bands has also been found in Y$_2$C \cite{Lu-JACS-2016, Inoshita-PRX-2014} (also a layered structure in anti-CaCl$_2$ type). However, the anionic electrons in $C$2/$m$-Rb$_3$O can also pass through the layers due to the presence of intralayer holes, while the excess electrons are only confined between the layers in Y$_2$C. To obtain more insights, we plotted the decomposed charge density for each band. As shown in Fig. \ref{band}e-h, the results suggest that the electrons accumulate around the center between two adjacent vacancies in Band 1, while around the vacancies in Band 2. If we zoom in the flat regions on both bands, they correspond to electrons forming 1D channel around the vacancies and 2D gas between the layers, respectively. Thus, we can obtain a qualitative understanding on the distribution of anionic electrons by bands. As we will discuss in the following section, these two bands and the overall distribution of anionic electrons can be effectively changed by external forces and chemical substitution.

\begin{figure*}
\epsfig{file=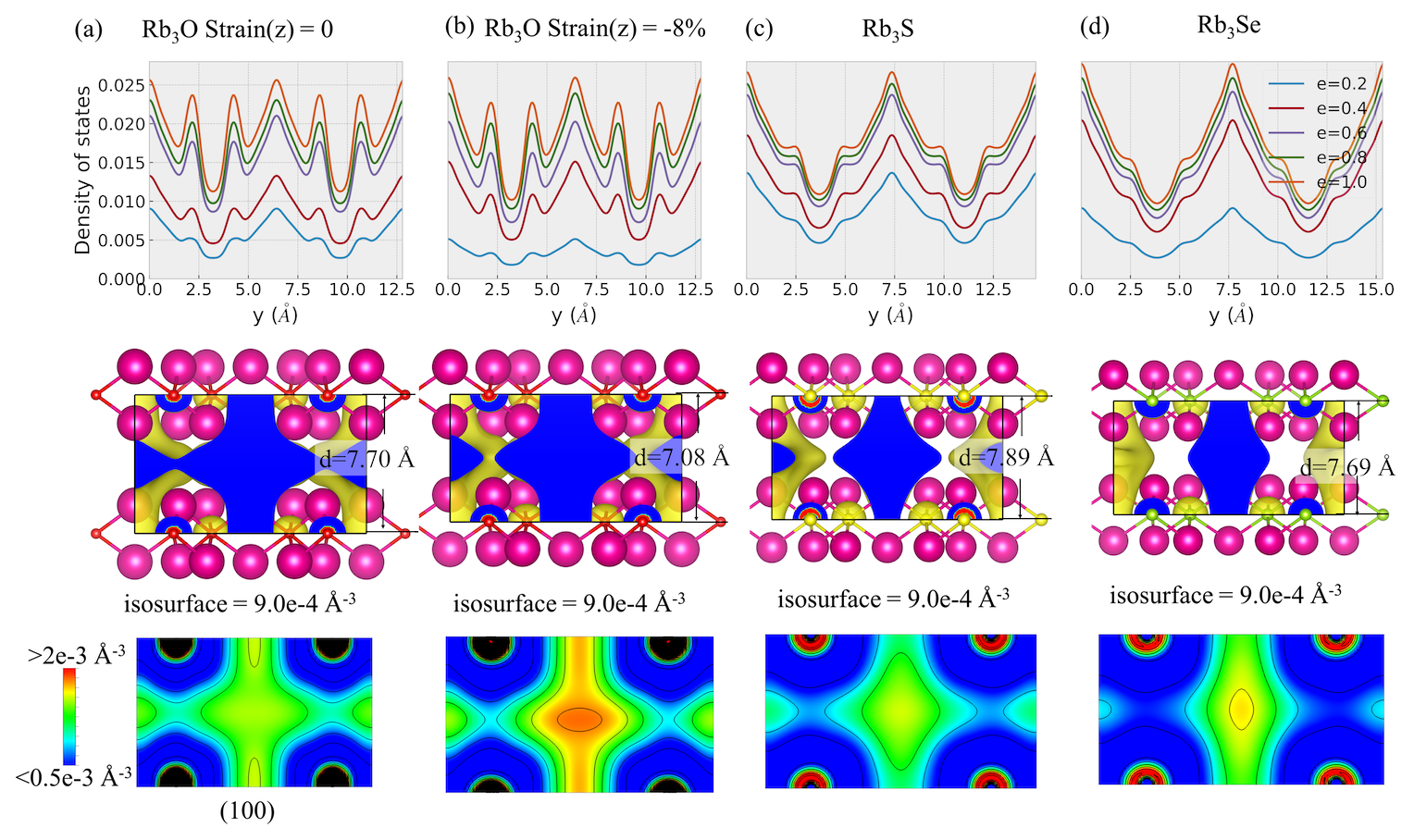, width=0.95\textwidth}
\caption{\label{strainandchem} {The influence of strain and chemical substitution. The calculated partial charge density distribution of (a) pristine $C$2/m-Rb$_3$O, (b) $C$2/m-Rb$_3$O under -8\% uniaxial strains along z direction, (c) $C$2/$m$-Rb$_3$S and (d) $C$2/$m$-Rb$_3$Se. Upper panel: the integrated density of states at different energies (0.2, 0.4, 0.6, 0.8 and 1.0 eV) along y-axis. Middle panel: The isosurfaces at 0.0009 {\AA$^{-3}$} for -1.0$ \leq E$-$E_{\textrm{Fermi}} \leq$0 eV. Lower panel: the sliced partial charge maps (-1.0$\leq E$-$E_{\textrm{Fermi}} \leq$0 eV) at (100) plane of the conventional cell. The magenta, yellow, green and red spheres denote Rb, S, Se and O atoms, respectively.}}
\end{figure*}

\subsection{Tuning the Interstitial Electrons}
Though $C$2/$m$-Rb$_3$O has a layered structure, the presence of O-vacancies makes it intrinsically different from the well-known Ca$_2$N type materials. According to the electrostatics, the excess electrons tend to be pushed away from the cation sites to the crystal cavities. In $C$2/$m$-Rb$_3$O, the excess electrons are trapped by two types of crystal cavities: i) the intralayer O vacancies; ii) the interlayer space. If there exist ways to leverage their relative electrostatic potential distribution, the trapped electrons might rearrange themselves into different distribution and even different topology. In the following sections we will demonstrate that such redistribution can be triggered by elastic strain and chemical substitution.

\textit{Elastic strain}. Strain engineering has been widely used to induce phase transition in a wide range of materials \cite{Li-MRS-2014}. Given the layered nature of $C$2/$m$-Rb$_3$O, it is intuitive to modify the size of the interlayer spacing by applying strain along the c-axis. Specifically, we applied up to 8\% compression along the c-axis. Our phonon calculations suggested that the structure after the deformation remains dynamically stable (see Fig. S6 \cite{SI}), indicating that Rb$_3$O is an ultra-compressible material which can sustain large elastic strain, like other 2D materials. Since the distribution of anionic interstitial electrons is a function of 3D space and energy, the interpretation of results can be challenging. In order to characterize them in a more comprehensive manner, we tracked the changes by calculating several properties, (i) the isosurface plots at different values; (ii) the integrated PDOS along three main crystal direction; (iii) the contour maps at some primary planes. Fig. \ref{strainandchem} shows the results for some important crystal directions. While the choice of isosurface value yields somewhat ambiguous pictures (since this is truly isovalue dependent), the integrated PDOS and contour maps tell more insights into the distribution over the entire parameter space. Compared with Rb$_3$O without strain, we observed the distribution of interstitial electrons has a strong localization in the (100) plane, see Fig. \ref{strainandchem}b partial charge map. This is also evidenced by the sharpening of the peak in the middle of the path along the y-axis (Fig. \ref{strainandchem}b). The changes on interlayer spacing also substantially changes the distribution in energy space. Clearly, the integrated PDOS values in Rb$_3$O under strain has more population in the energy range between 0.2 and 0.4 eV relative to $E_\textrm{Fermi}$. As shown in Fig. S8 \cite{SI}, we found that the accumulation of PDOS from -0.5 eV to 0 eV is even more obvious.

\textit{Chemical substitution}. Clearly, the topology of interstitial electrons largely depends on the arrangement of vacancies and the interlayer distances. The chemical substitution is another common approach to adjust these parameters. We therefore checked the possibility of Rb$_3$S and Rb$_3$Se based on the same structural motif. Both of them were calculated to be unstable with respect to the convex hull (0.06 eV/atom for Rb$_3$S and 0.08 eV/atom for Rb$_3$Se at the level of optPBE). Rb$_3$S is dynamically stable at both PBE and optPBE level, while Rb$_3$Se is dynamically stable at PBE level but unstable at optPBE level based on the phonon calculations (see Fig. S7 \cite{SI}). Without considering the vdW force, the interlayer distance monotonically increases from O, S to Se, while considering the vdW interaction, the interlayer distance first increased at S but then dropped at Se due to strong interlayer Se-Se vdW attraction. It suggests that vdW interaction is crucial to describe the correct geometry. Therefore, we discuss the results based on the geometries taking into account of vdW correction. As shown in Fig. \ref{strainandchem}c-d, the anionic electrons in both Rb$_3$S and Rb$_3$Se appear to lose the connectivity along the y-axis from the isosurface plots (note that we set the same isosurface value 9.0e-4/{\AA}$^3$ for all systems under consideration). This is also supported by the results from contour plots and integrated PDOS. In the integrated PDOS plot, both Rb$_3$S and Rb$_3$Se (Fig. \ref{strainandchem}c-d) show nearly single peak character along the y-axis, as opposed to the multiple peaks in Rb$_3$O (Fig. \ref{strainandchem}a). Although there still exist some minor percentages of DOS along the y-axis, it is clear that both Rb$_3$S and Rb$_3$Se tend to behave more like 2D electrides, suggesting that chemical substitution is a more effective approach to tune the anionic electrons than strain engineering. This is not surprising, since it results in a more uniform expansion in intralayer spacing by replacing O with S/Se. In the meantime, both S and Se are less electronegative than O. Therefore, the excess electrons tend to accumulate more in the vacancy sites. Last, we emphasize that the synthesis of pure Rb$_3$S and Rb$_3$Se is thermodynamically unfavorable, but the partial substitution on O sites of Rb$_3$O is possible. Therefore, the topology of $C$2/$m$-Rb$_3$O becomes rather flexible via controlling the concentration of the substituted S/Se. 

\begin{figure}
\epsfig{file=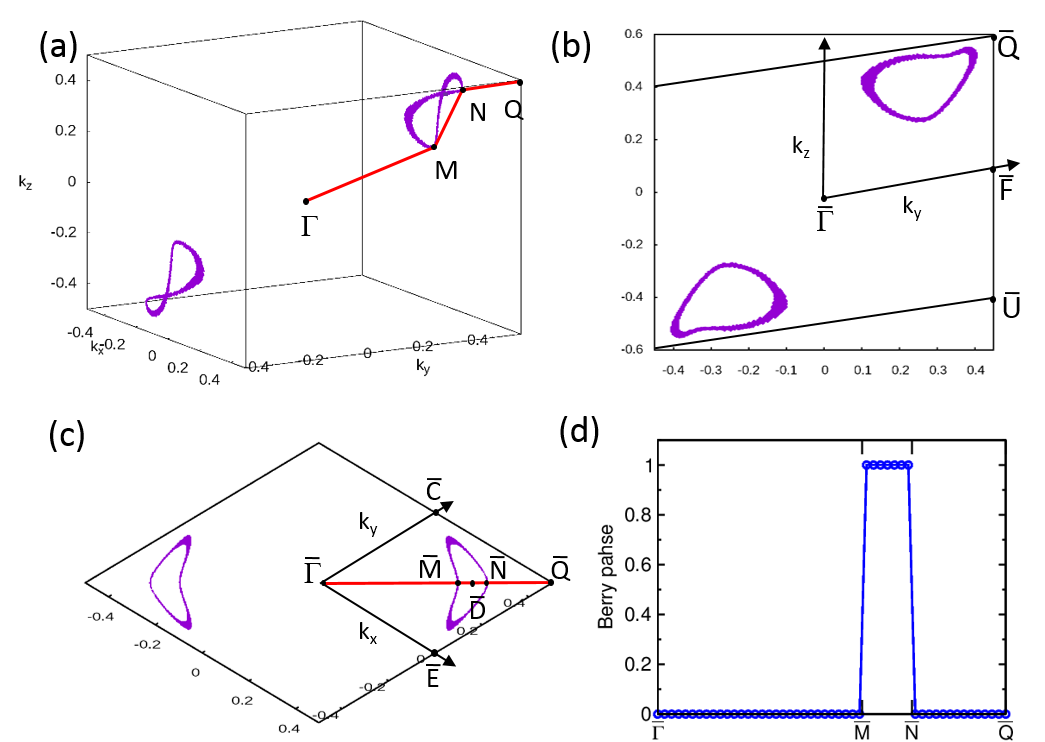, width=0.45\textwidth}
\caption{\label{cell}{(a) Two degenerate nodal rings in the reciprocal primitive cell (RPC), which includes four characteristic points ($\Gamma$ (0.0, 0.0, 0.0), $M$ (0.5, 0.5, 0.5), $X_1$ (0.2971, 0.2971, 0.2174) and $X_2$ (0.3623, 0.3623, 0.4493)). Among them, $X_1$ and $X_2$ are two points on the nodal rings. (b) Projected RPC along [100] direction. (c) Projected RPC along (001) direction. (d) The variation of the Berry phase along path $\overline{\Gamma}$-$\overline{X_1}$-$\overline{X_2}$-$\overline{M}$ of the (001) projection.}}
\end{figure}

\begin{figure}
\epsfig{file=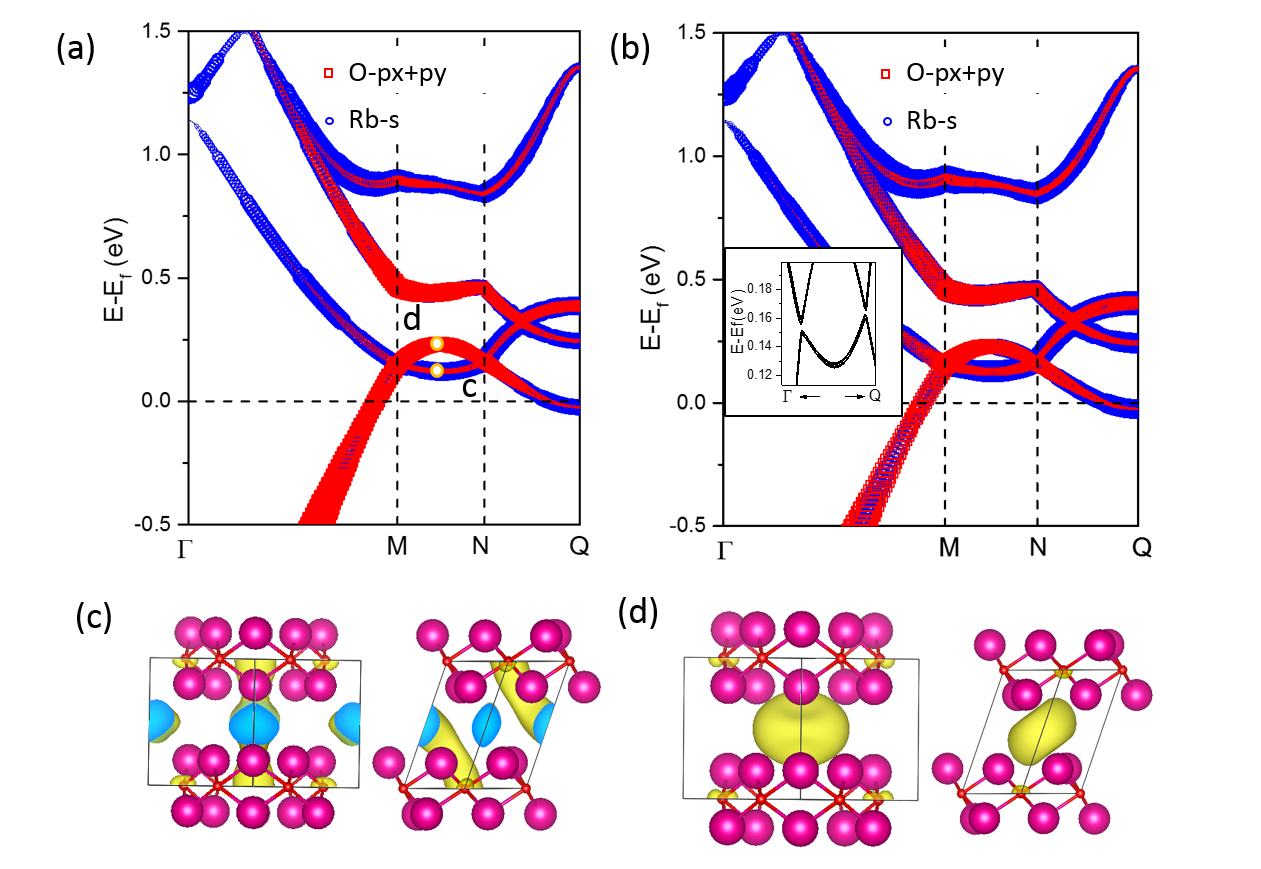, width=0.45\textwidth}
\caption{\label{BandStructure}{Electronic band structures of Rb$_{3}$O without (a) and with (b) the SOC. The inset in (b) shows the enlarged band structure around $X_1$ and $X_2$ points with SOC. (c) and (d) show the wave functions of two eigenstates marked as c and d in the band structure (left: front view, right: side view).}}
\end{figure}

\subsection{Non-trivial electronic topological properties}
It was recently proposed that the so-called unique floating electrons, which occupy the interstitial space, in the electrides are favorable for achieving band inversions needed for topological phase transition in the electronic states \cite{Hirayama-PRX-2018}. Indeed, with this mechanism a few known electride materials, including Y$_2$C \cite{Hirayama-PRX-2018,Huang-NanoL-2018}, Ca$_3$Pb \cite{Ca3Pb-2018}, Cs$_3$O \cite{Park-PRL-2018} and Sr$_2$Bi \cite{Hirayama-PRX-2018}, were identified to be topological non-trivial. As a result, we have paid attention to the currently predicted electride material of Rb$_3$O. From our currently obtained electronic band structure without the spin-orbit coupling (SOC) in Fig.~\ref{band}a, it seems to show the nearly linear band crossings between the energy-lowest conduction band (CB) and the energy-highest valence band (VB) around the Fermi level at the high-symmetry points, \emph{L} and \emph{Z}, in the BZ. After a careful double checking on these two crossings, they indeed exhibit small opening gap of 22 meV and 24 meV at the \emph{L} and \emph{Z} points, respectively. It is well-known that, when the time reversal symmetry and the space inversion symmetry coexists in a lattice system, the energy inverted bands with opposite parities have to cross a closed Dirac nodal line if its SOC effect is ignored. This would be exactly what happens for Rb$_{3}$O. Since \emph{L} and \emph{Z} are not strictly band crossing points, we have performed an extended search in the whole lattice momentum space with a dense 120$\times$120$\times$120$~\vec{k}$-mesh to identify whether or not the nearly linear band crossings exist between the CB and VB bands in the BZ. Interestingly, we have found that two Dirac nodal lines (DNLs) exist, as shown in Fig.~\ref{cell}a. It is important to note that these DNLs cannot be easily found because they do not belong to any high-symmetry path. To elucidate the nature of the band inversion for these DNLs as discussed in previous literature~\cite{Li-PRL-2016,Wang-PRB-2012,Cheng-PRB-2014}, we labeled two special points $X_1$ and $X_2$ on the DNLs in the BZ (see Fig.~\ref{cell}a and \ref{cell}c) and then plotted the corresponding electronic band structures along the $\Gamma$-$X_1$-$X_2$-$M$ path without SOC in Fig.~\ref{BandStructure}a and with SOC in Fig.~\ref{BandStructure}b. Without the SOC inclusion there are two strict band crossings (corresponding to Dirac cones) at about 0.2 eV above the Fermi level along the $X_1$-$X_2$ path, whereas the SOC inclusion results in small openings of these two crossing points as illustrated in the inset of Fig.~\ref{BandStructure}b. In fact, these band crossings do not occur only at these two points and, instead, they continuously cross each other to form the three-dimensional (3D) snake-like wrapped DNLs. This kind of special DNLs have been reported in some recent work~\cite{Huang-PRB-2016}, that are typically different from those in Be \cite{Li-PRL-2016}, CaP$_3$ \cite{Xu-PRB-2017} and Ca$_2$P$_3$ \cite{Chan-PRB-2016} in which nodal lines lie in a 2D plane. To elucidate the mechanism of the formation of the DNLs in Rb$_{3}$O, we have analyzed the band inversion along the defined $X_1$-$X_2$ path, which is induced by the Rb-$s$-like and O-$p_x+p_y$-like orbitals (as marked by in Fig.~\ref{BandStructure}a), because they possess opposite positive and negative parities. Accordingly, we have plotted the decomposed charge densities in Fig.~\ref{BandStructure}c-d at the selected middle k-point between $X_1$ and $X_2$ in Fig.~\ref{BandStructure}a. It can be seen that the charges of the inverted bands at the selected positions (d and c as marked in Fig.~\ref{BandStructure}a) are localized in the crystal interstitial sites, which is a typical fingerprint for an electride. In addition, as already observed in Fig.~\ref{BandStructure}b the DNLs will split into tiny gap once the SOC effect is switched on. The tiny gap is expected since both Rb and O have relatively light masses and the resulting SOC effect is likely to be negligible.

\begin{figure*}
\epsfig{file=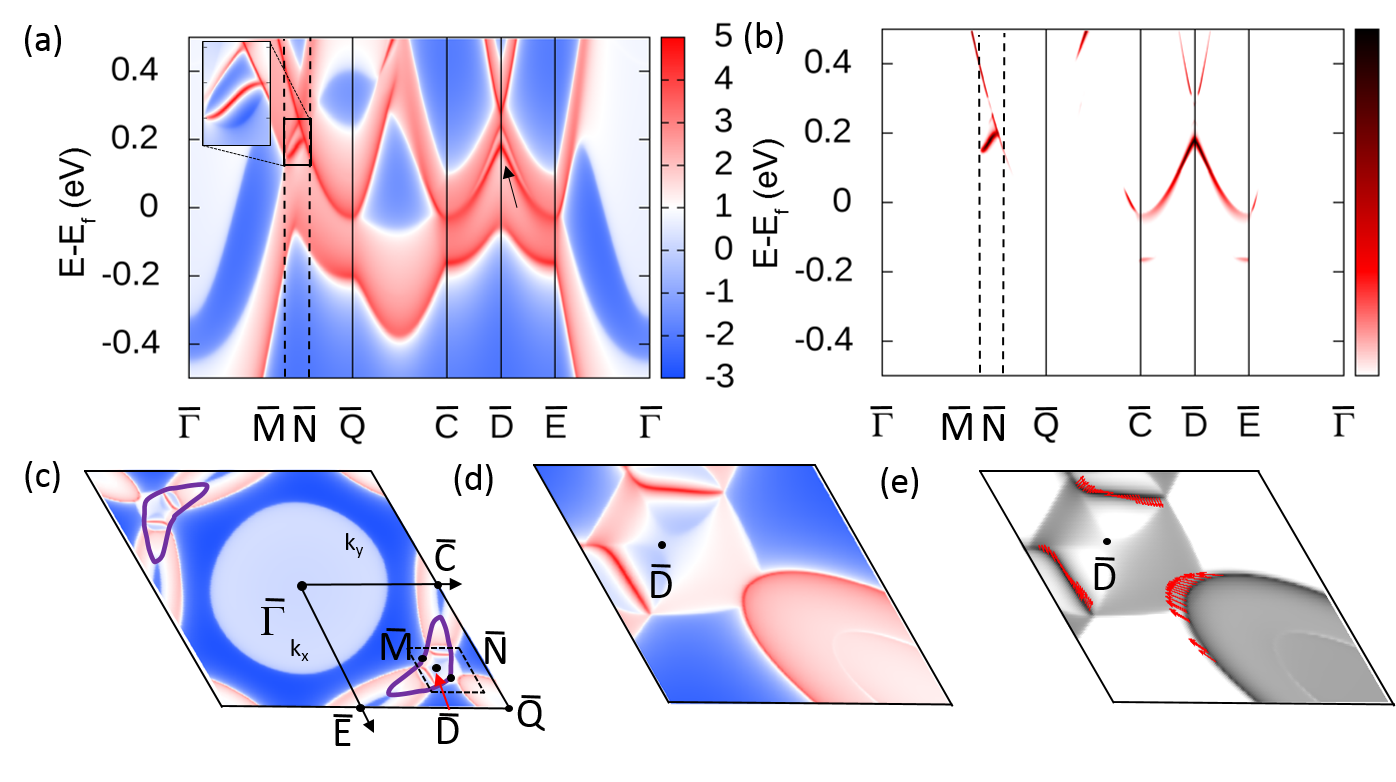, width=0.80\textwidth}
\caption{\label{Surfacestate}{(a) Projected surface band spectrum of the (001) surface of Rb$_{3}$O, where $\overline{C}$ and $\overline{E}$ correspond to (0.5, 0) and (0.5, 0), and $\overline{D}$ is the middle point between $\overline{X_1}$ and $\overline{X_2}$. (b) The difference between surface and bulk band spectrum for the (001) surface of Rb$_{3}$O. (c) The Fermi surfaces for the (001) surface at $E_{\textrm{Fermi}}$ + 135 meV, (d) the zoomed-in image of the plot around $\overline{D}$, and (f) the in-plane spin texture of the surface state at $E_{\textrm{Fermi}}$ + 135 meV.}}
\end{figure*}

To further identify the non-trivial topological nature of these DNLs in Rb$_{3}$O, we have investigated the Berry phase in the \emph{$k_{x}$}-\emph{$k_{y}$} plane of Rb$_{3}$O. The Berry phase \cite{Chan-PRB-2016,Huang-PRB-2016,Bian-PRB-2016} in the \emph{k$_{x}$}-\emph{k$_{y}$} plane can be expressed as follows,

\begin{equation}\label{berry}\emph{P}(k_{x},k_{y})=
-i \sum_{E_{j}<E_{f}}\int_{-\pi}^{\pi}\left\langle
u_{j}(\mathbf{k})|\partial _{k_{z}} |u_{j}(\mathbf{k})\right \rangle dk_{z},
\end{equation}

where $u_{j}(\mathbf{k})$ is the periodic part of the bulk Bl\"och wave function in the \emph{j}th band. The sum is over all occupied Bl\"och eigenstates $|u_{j}(\mathbf{k})\rangle$ of the Hamiltonian \cite{Chan-PRB-2016}. As proposed in Ref.~\cite{Chan-PRB-2016}, the Berry phase $\emph{P}(k_{x},k_{y})$ with the SOC inclusion is quantized in unit of $\pi$, and it is related to the charge at the end of the 1D system with fixing $k_{x}$ or $k_{y}$. The Berry phase is defined modulo 2$\pi$, due to the fact that the large gauge transformation of the wave functions can change it by 2$\pi$. If $\emph{P}(k_{x}, k_{y})$ = $\pi$ mod 2$\pi$, there exists an odd number of the drumhead surface states at ($k_{x}$, $k_{y}$) in the surface BZ~\cite{Chan-PRB-2016}. Following this method, we have calculated the Berry phase along the $\overline{\Gamma}$-$\overline{X_1}$-$\overline{X_2}$-$\overline{M}$ path of the (001) surface BZ (Fig.~\ref{cell}d). It can be seen that the value of the Berry phase suddenly steps at the $\overline{X_1}$ and $\overline{X_2}$ points, indicating that DNLs crossing $X_1$ and $X_2$ points are topologically non-trivial. In particular, the Berry phase equals $\pi$ for all ($k_{x}$, $k_{y}$) points inside the projected DNL on the (001) surface, whereas the corresponding values are zero for all ($k_{x}$, $k_{y}$) points outside the projected DNLs. This fact implies that the non-trivial topological surface states certainly occur within the (001) projected nodal rings of the bulk DNLs.

Now, let us focus on the topologically protected non-trivial surface electronic states. We have first compiled the (001) surface electronic band structure in Fig.~\ref{Surfacestate}a along the selected \emph{k}-path as defined in Fig.~\ref{cell}c in order to visualize the DNL-induced drumhead-like non-trivial surface states. As expected, two apparent non-trivial surface states can be clearly visualized, as shown by the highest color density in the ranges from $\overline{X_1}$ to $\overline{X_2}$ path and from $\overline{C}$ to $\overline{D}$ to $\overline{E}$. The former whole range and the latter partial range cross the defined drumhead-like region projected by the bulk DNL onto the (001) surface. These non-trivial surface states are also evidenced by the difference between the surface electronic bands and the (001) projected states from the bulk electronic bands (Fig. \ref{Surfacestate}b). Furthermore, in order to see the evolution of drumhead-like non-trivial surface states by varying the energy, we have calculated the Fermi surfaces and the spin texture of the (001) surface. We have selected the energy of 135 meV above the Fermi level and plot the Fermi surface in Fig.~\ref{Surfacestate}c-d. At this energy, some isolated bright states distinguishable from the bulk states form surrounding the point $\overline{D}$ and nearly along the $\overline{X_1}$-$\overline{X_2}$ path, which certainly correspond to the drumhead-like non-trivial surface states induced by the bulk DNLs. Interestingly, we have found that the topological surface states are spin-polarized with helical spin textures (Fig. \ref{Surfacestate}e), which is originated from unique spin-momentum locking property in topological materials. It is noteworthy that such non-trivial surface states can be also observed in the (100) surface band structure (see Appendix \ref{app1}).

The identification of topological electrides requires the analysis of both topological and electride properties. The automated classification of topological materials along high symmetry point (HSP) or high symmetry line (HSL) has been enabled recently \cite{zhang2018catalogue, vergniory2018high, tang2018towards}. Similarly, two recent works reported the high-throughput identification of inorganic electrides from the entire experimental materials database \cite{burton2018high, zhu2018inorganic}. Comparing both the topological \cite{zhang2018catalogue} and electride \cite{zhu2018inorganic} materials databases, we found many materials belong to both catalogues \cite{zhu2018inorganic}. Indeed, all the recently reported topological electrides \cite{Hirayama-PRX-2018,Huang-NanoL-2018,Ca3Pb-2018,Park-PRL-2018} have been found and classified to either HSP (Ba$_3$N, Cs$_3$O) or HSL (Ca$_3$Pb, Y$_2$C, Sr$_2$Bi) \cite{zhu2018inorganic}. However, Rb$_3$O cannot be identified from such schemes. First, this structure is entirely from first-principles prediction. Second, the band inversion in Rb$_3$O occurs at the general electronic k momentum (neither HSP nor HSL), which is beyond the current capability of topological material identification \cite{zhang2018catalogue}. We anticipate that Rb$_3$O represents a very rare class of topological electrides which have not been explored fully. The accumulation of more examples would be useful to understand better the relation between interstitial electrons and topological properties.

\subsection{Extension to Other Binaries} 
In the recent years, high-throughput computational screening has been playing an increasingly important role in the discovery of new electrides \cite{Inoshita-PRX-2014, Ming-JACS-2016, Zhang-PRX-2017, Tada-IC-2014}. Based on the Ca$_2$N prototype, a series of 2D electrides have been proposed \cite{Tada-IC-2014}. For example, Y$_2$C has been synthesized in the laboratory since 1960s \cite{atoji1969crystal, maehlen2003structural}. However, it was identified to be electrides by calculation only recently \cite{Inoshita-PRX-2014}. On the other hand, the recent advances in crystal structure prediction made it possible to search for new materials with unknown structure prototypes \cite{Wang-JACS-2017, Ming-JACS-2016, Zhang-PRX-2017}. For example, Zhang et al \cite{Zhang-PRX-2017} proposed an inverse design strategy to search for inorganic electrides. Using the interstitial electron's ELF as the optimization target, they identified a number of potential electrides A$_2$B and AB compounds \cite{Zhang-PRX-2017}. Furthermore, Wang et al \cite{Wang-JACS-2017} pointed out that one had to consider the possibility of competing phases with different stoichiometries in the crystal structure search. By applying the variable compositional structure search of Sr-P system, they found two new stable electrides Sr$_8$P$_5$ and Sr$_5$P$_3$. However, such first-principles CSP is computational expansive, and thus the applications were limited to a small set of chemical systems.

\begin{figure*}
\epsfig{file=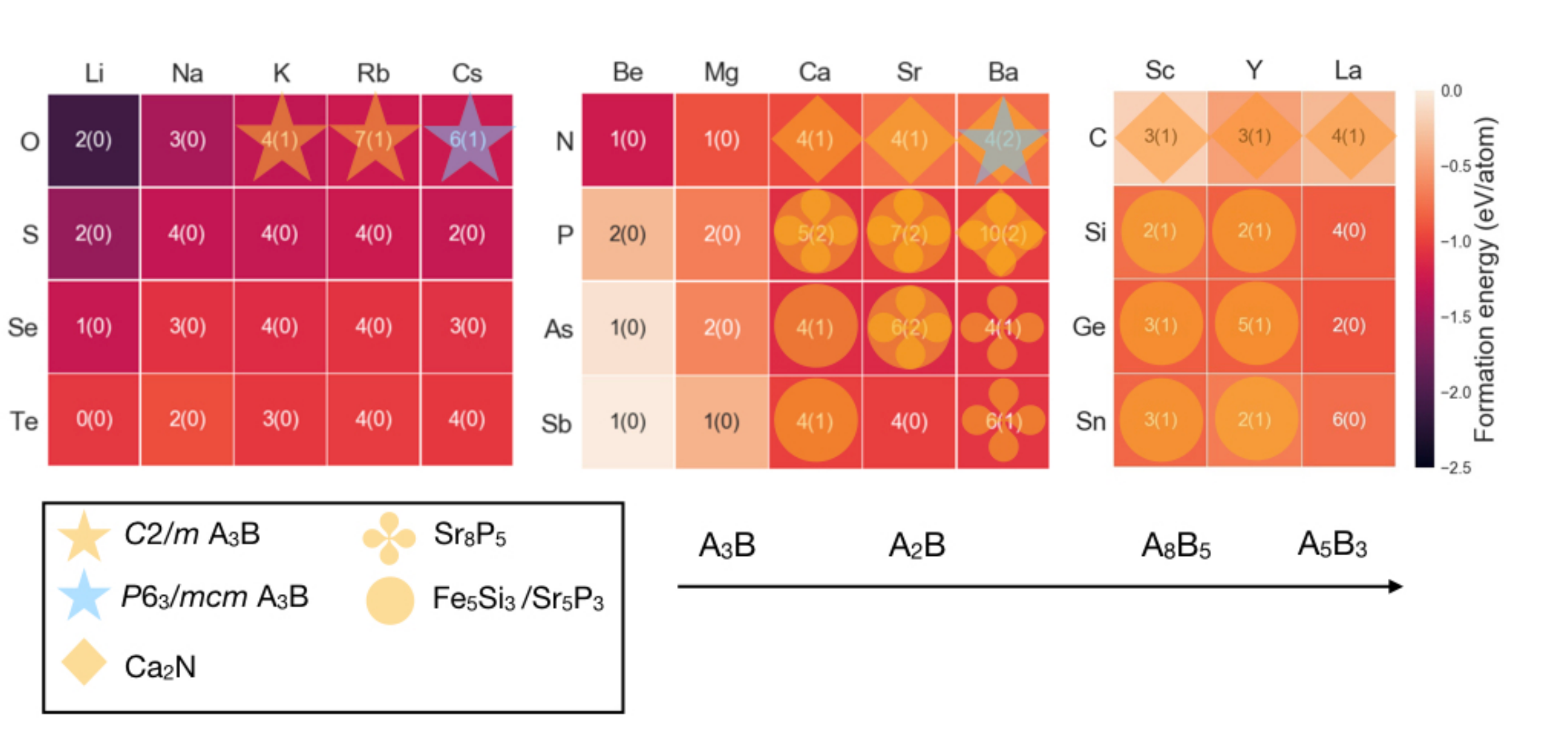, width=0.85\textwidth}
\caption{\label{ht} {The heat map of group IA-VIA, IIA-VA and IIIB-IVA binaries. Each box of binary is colored by the compounds with most negative formation energy in the system, followed by the number of stable compounds and the number of stable electrides (in parenthesis).}}
\end{figure*}

In order to gain a more complete information about phase stability, we combined the advantages of both data mining and first-principles CSP methods. We selected the best ten structures by fitness ranking from CSP runs on both Rb-O and Cs-O, and then applied chemical substitution to other binaries in group IA-VIA, IIA-VA and IIIB-IVA. We also included the prototypical structures of several well-known electrides, such as A$_2$B (anti-CaCl$_2$ type \cite{Lee-Nature-2013}), A$_5$B$_3$ (Fe$_5 $Si$_3$ and Sr$_5$P$_3$ types) \cite{Lu-JACS-2016}, A$_8$B$_5$ (Sr$_8$P$_5$ type) \cite{Wang-JACS-2017} in our study. Taking advantage of the data available in the Materials Project \cite{MP-2013}, we computed the final convex hull diagram for each binary system under investigation at PBE level. The complete information can be found in Appendix B. Fig. \ref{ht} summarizes the distribution of stable electrides in these systems. Our analysis identified several other binary systems including alkaline metal oxides and alkaline earth metal nitrides in which the A$_3$B compounds are thermodynamically stable. It is obvious that these compounds are favored in the systems characterized by large ratio of cation/anion sizes and large electronegativity difference. For instance, A$_3$B are unstable in Li-O and Na-O. With the increase in the cation size, the $C$2/$m$ K$_3$O with a larger cation size becomes stable at PBE level, but marginally stable at optPBE level (only 9 meV/atom above the convex hull). Thus, $C$2/$m$ K$_3$O  is also likely to be synthesized in experiment. The further increase of cation size leads to that the 1D anti-$\beta$-TiCl$_3$ type structure with a hexagonal lattice ($P$6$_3/mcm$) becomes more favorable. The corresponding stable structures (Cs$_3$O and Ba$_3$N) are also electrides, according to recent work by Park \textit{et. al} \cite{Park-PRL-2018}. Both anti-$\beta$-TiCl$_3$ type Cs$_3$O and Ba$_3$N have been synthesized experimentally in the past \cite{Tsai-JPC-1956, Steinbrenner-1998}, while the discovery of Rb$_3$O and K$_3$O in this work provides an example of unknown electrides entirely predicted from first-principles. More importantly, the monoclinic $C$2/$m$ is so far the only discovered phase with tunable topology of anionic electrons. The phase stability is also associated with electronegativity. Extending to group IIIB-IVA, we failed to find any stable A$_3$B electrides. Instead, the electrides with higher concentrations of anions (A$_2$B and A$_5$B$_3$) tend to become thermodynamics favorable in these systems (Fig. \ref{ht}). 

\section{DISCUSSION} 

\begin{figure*}
\epsfig{file=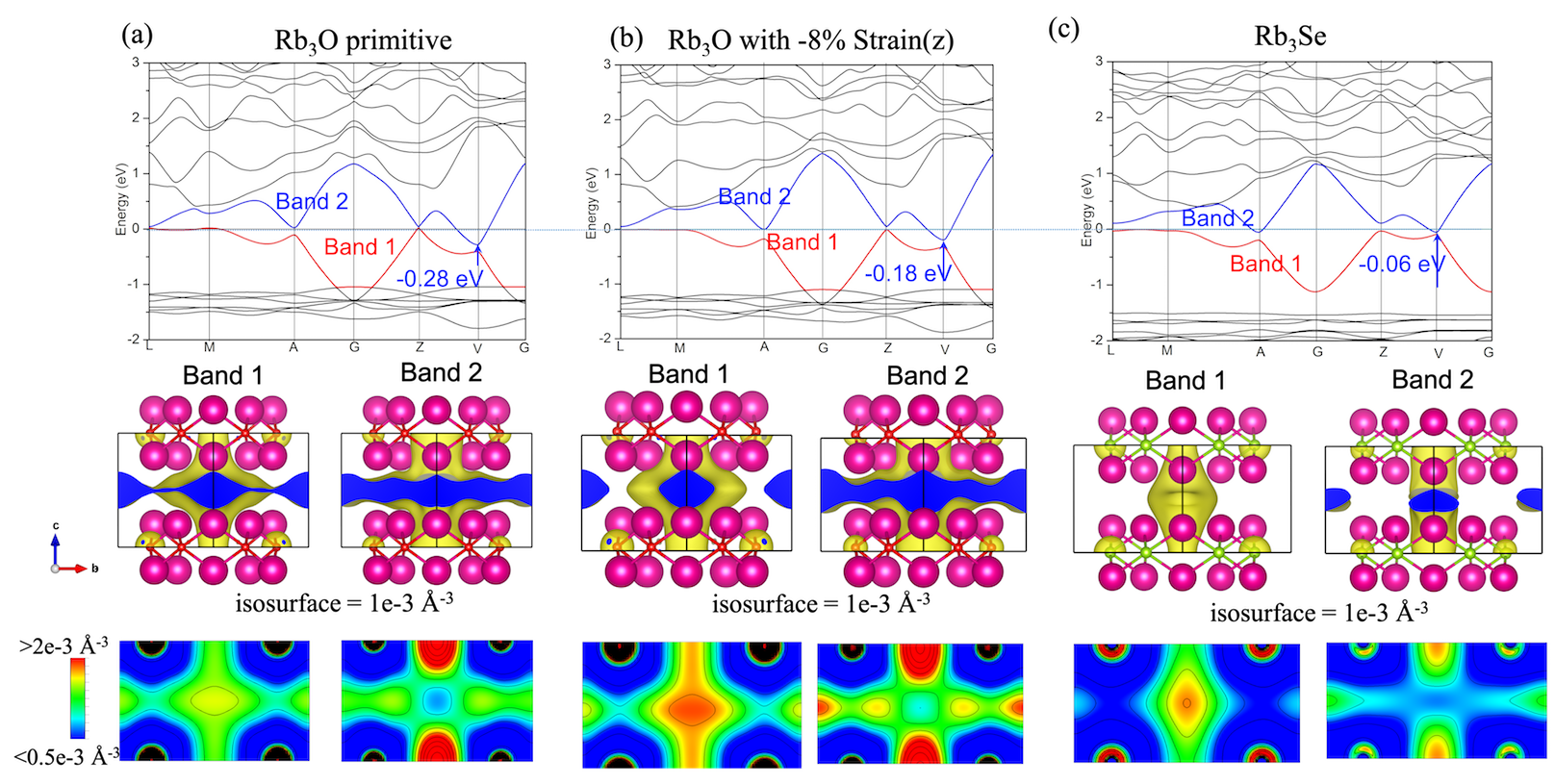, width=0.85\textwidth}
\caption{\label{mechanism} {The electronic band structure (upper), the isosurface (middle) and map (lower) of density distribution of interstitial Band 1 and Band 2 of $C$2/$m$ Rb$_3$O phase at zero strain (a), 8\% strain (b) and Rb$_3$Se (c). The interstitial bands are highlighted with red and blue lines. The blue arrows indicate the gradual upper shift of band 2 towards $E_{\textrm{Fermi}}$. The large magenta spheres denote Rb atoms, the small red spheres denote O atoms and the small green spheres denote Se atoms.}}
\end{figure*}

The discovery of electrides has attracted a lot of attention recently as they are examples of exotic materials with anionic electrons and have many potential energy applications. While a number of electride systems with different topologies have been already reported in the literature, here we provide a first example of an electride material with flexible topologies of the interstitial electrons and discuss its relation with certain features of crystal packing. From the atomic point of view, the tunable electrides are quite different from other types of electrides. The candidate crystal structures are expected to contain multiple types of crystal cavities, such as 1D channels and 2D layered spacings in the same material. For example, the $C$2/$m$-Rb$_3$O electride studied in this work has both interlayer vacancies and interlayer spacing available to accommodate the interstitial electrons. Obviously, the multiple electron localization sites are crucial for changing the anionic electron's distribution. Though the electrides with multiple types of crystal voids have been reported in the past \cite{wagner1995cs+,dale2014density,wu2017tiered}, our present work shows that the external stain and chemical substitution could breakdown the connection and thus lead to a substantial change in the anionic electron's topology.

How the electron topology can be switched in these types of electrides? The distribution of the excess electrons in the crystal is dictated by many factors, including Coulomb repulsion, kinetic energy and Pauli exclusion, which can be altered in a number of ways. For example, varying the interlayer spacing with mechanical strain can induce the redistribution of interstitial electrons. Without compressive strain, there exist two bands crossing $E_{\textrm{Fermi}}$. Upon compression, Band 2 undergoes an upper shift from -0.28 eV to -0.18 eV, while more anionic electrons tend to occupy on Band 1. Alternatively, chemical substitution can change the electron transfers in the structure, as well as the interlayer and interlayer spacing, thus it may lead to a more profound change. As shown in Fig. \ref{mechanism}, the continuing upper shift of Band 2 from Rb$_3$O to Rb$_3$Se eventually leads to a switch from 3D to 2D-like topology. Based on these observations, we conclude that the flexible nature of anionic electrons is due to the response of electronic bands against the subtle changes of local atomic environment. Our work here only explored two ways of adjusting excess electrons distribution through the structure modifications. However, other approaches, such as combined external electric or magnetic fields, may produce similar effects if the electrides contain different types of crystal cavities.

\section{CONCLUSION}
In summary, we identify a family of electrides with flexible topologies, by combining the evolutionary crystal structure prediction with high-throughput screening. These electrides are composed of 2D-layered Rb$_3$O motifs but have different stacking sequence and different number of O-vacancies. Due to the presence of two types of crystal cavities in layered Rb$_3$O, the interstitial electrons have viable responses to the subtle change of local environment, which may lead to substantially different distribution by mechanical strain or chemical substitution. We demonstrated that the key to design new electride materials with flexible anionic electron topology is to build the plausible crystal structures containing multiple types of crystal cavities and identify the ways to localize the electrons in these different cavities. Moreover, our calculation suggested that Rb$_3$O is a unique topological Dirc nodal line semimetal with a peculiar 3D-snake-like shape and band inversions occurring not along any high-symmetry path. We believe our investigation on Rb$_3$O and other binary systems offers new strategies for search and design of other flexible electrides with combined topological properties and thus is beneficial to the future functional electride materials design.

\begin{acknowledgments}
Work at UNLV is supported by the National Nuclear Security Administration under the Stewardship Science Academic Alliances program through DOE Cooperative Agreement DE-NA0001982. Work in China was supported by the National Science Fund for Distinguished Young Scholars (No. 51725103) and NSFC fund (No. 51671193 and 21703004). J.-Y. Qu acknowledges the China scholarship council support (No.201706350087). We acknowledge the use of computing resources from XSEDE and Center for Functional Nanomaterials under contract no. DE-AC02-98CH10086. Support for T. Frolov was provided under the auspices of the U.S. Department of Energy by Lawrence Livermore National Laboratory under Contract DE-AC52-07NA27344.
\end{acknowledgments}

S.Z. and L.W contributed equally to this work.

\bibliography{reference}

\appendix
\section{Projected states for the (100) surface of Rb$_3$O}
\label{app1}
\begin{figure}
\epsfig{file=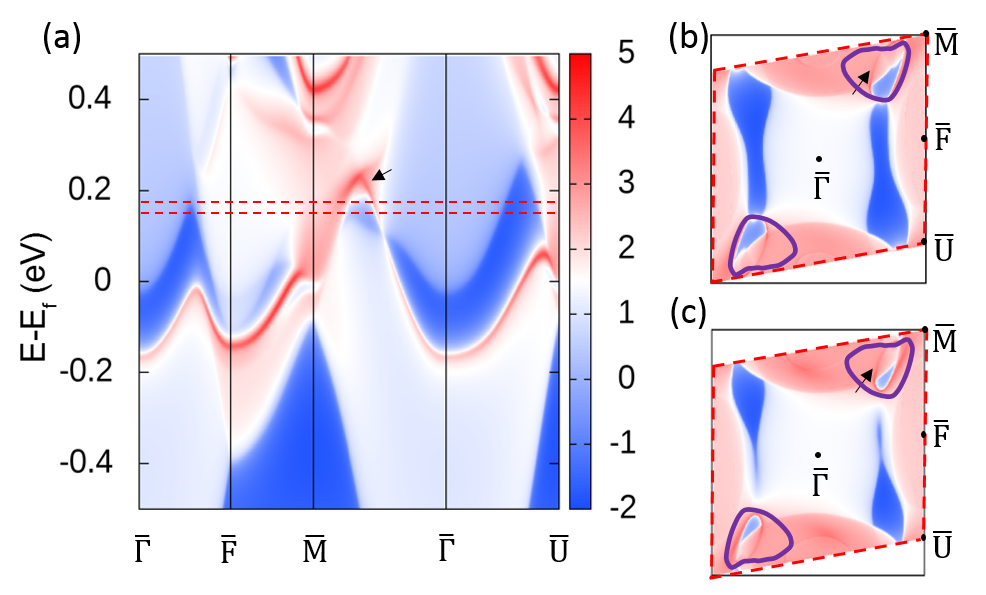, width=0.45\textwidth}
\caption{\label{Surfacestates}{(a) Projected surface spectrum of for the (100) surface of Rb$_{3}$O, where $\overline{M}$, $\overline{F}$ and $\overline{U}$ correspond to (0.5, 0.5), (0.5, 0), (0.5, -0.5), respectively. The Fermi surfaces for the (100) surface at $E_{\textrm{Fermi}}$ + 150 meV (b) and at $E_{\textrm{Fermi}}$ + 180 meV (c). The black dash ring is the projection in (100) surface of nodal ring.}}
\end{figure}

The electronic band structure and the Fermi surfaces are shown in Fig. \ref{Surfacestates}a. In the path along $\overline{\Gamma}$-$\overline{F}$-$\overline{M}$, there are some bright surface states. In comparison with the region at which the drumhead-like non-trivial surface states exist, it has been recognized that these bright surface states outside the projected nodal ring are apparently trivial surface states. Certainly, we have also found some drumhead-like non-trivial surface states as marked by the black arrow, which are within the projected nodal ring. These non-trivial surface states in an apparent ellipse-like shape can be also observed in the Fermi surfaces at $E_{\textrm{Fermi}}$ + 150 meV and $E_{\textrm{Fermi}}$ + 180 meV in Fig. \ref{Surfacestates}b-c, respectively.
\section{Extended phase stability calculations on other binary systems}
\begin{figure*}
\epsfig{file=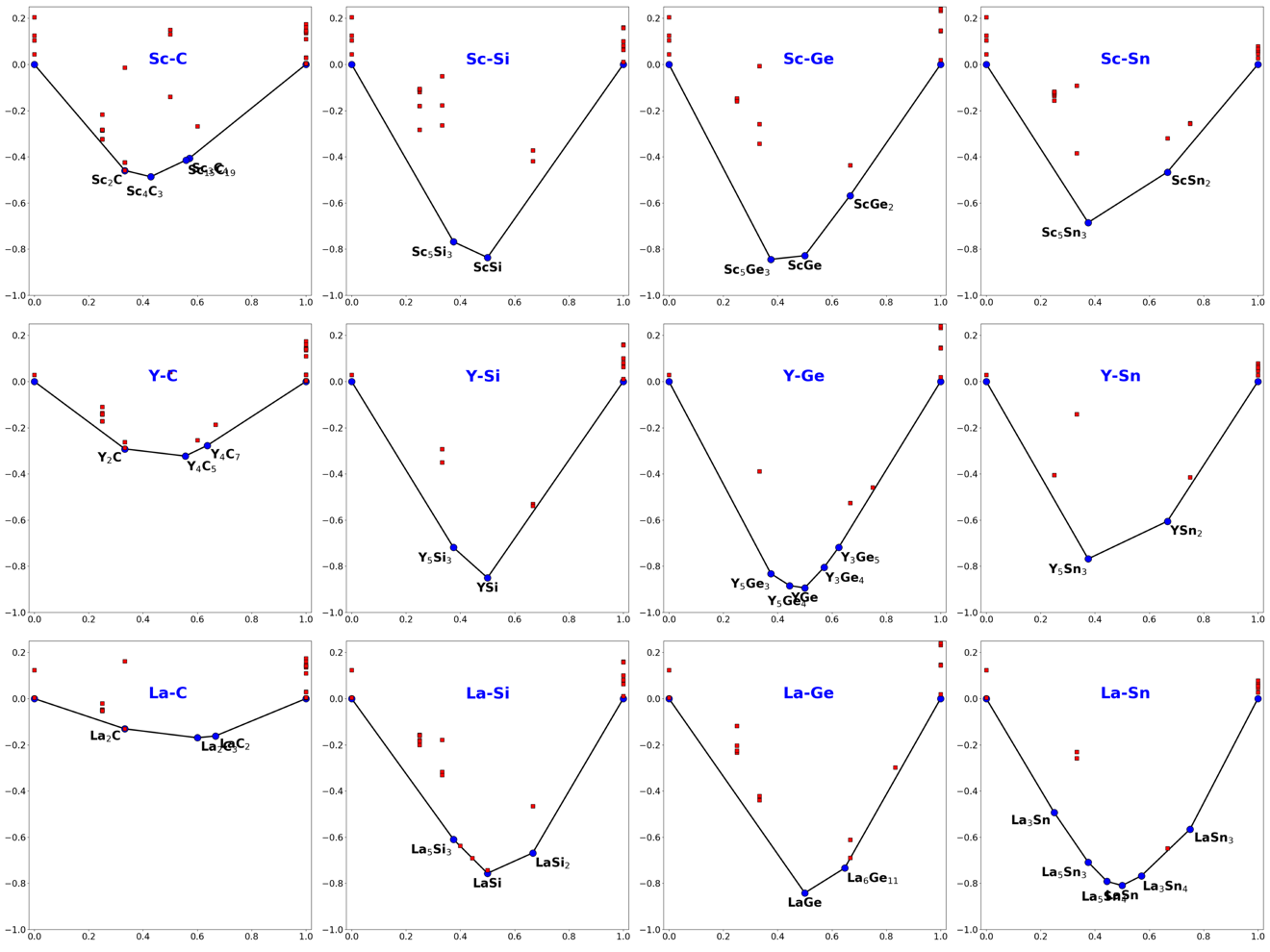, width=0.9\textwidth}
\caption{\label{ht1} {The computed convex hull diagrams of group IIA-VA binary systems at PBE level based with the same parameters setting used in Materials Project. In each diagram, the x-axis is composition ratio, and the y-axis is the formation enthalpy in eV/atom relative to the ground states of the corresponding elemental allotropes.}}
\end{figure*}

\begin{figure*}
\epsfig{file=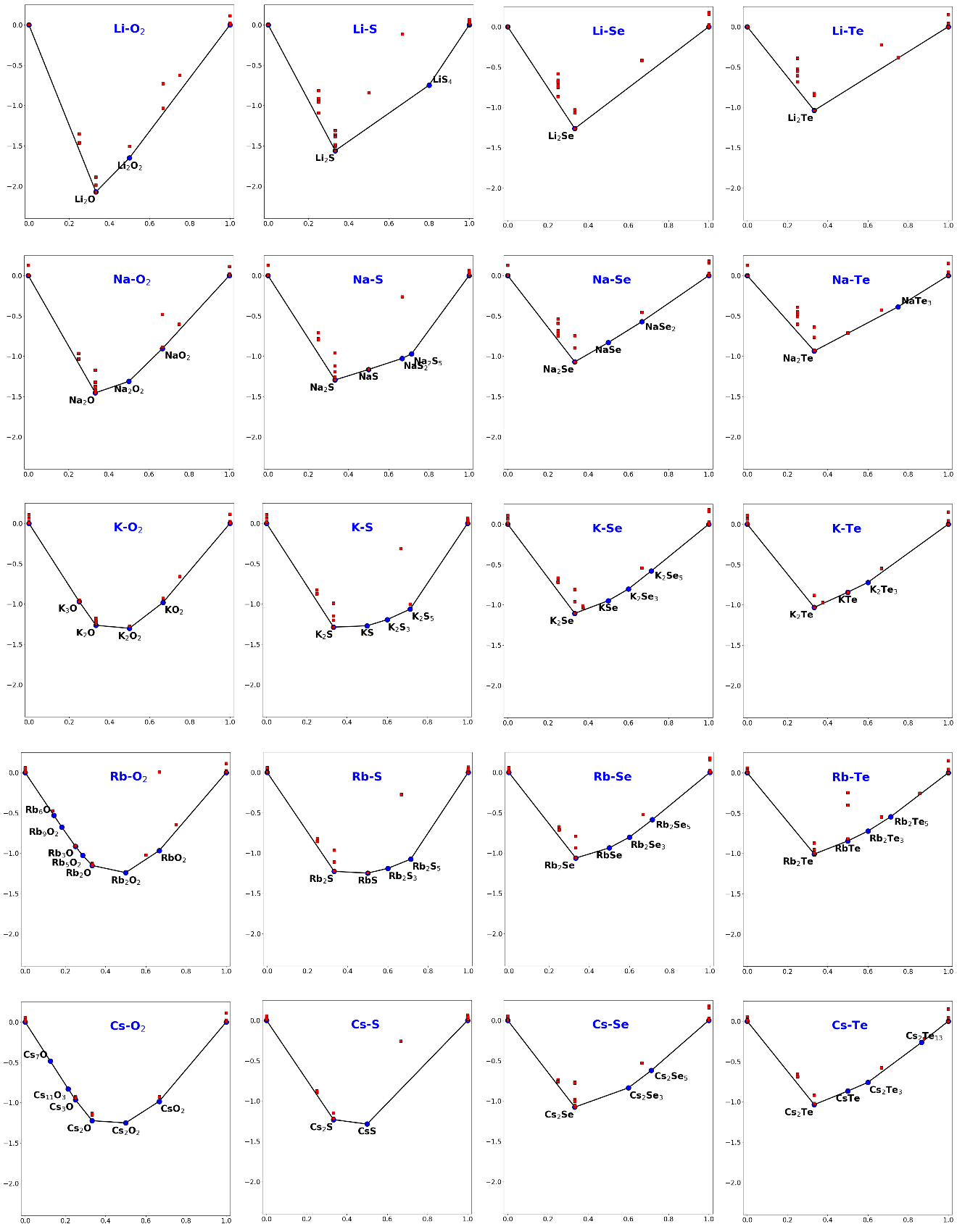, width=0.9\textwidth}
\caption{\label{ht2} {The computed convex hull diagrams of group IA-VIA binary systems at PBE level based with the same parameters setting used in Materials Project. In each diagram, the x-axis is composition ratio, and the y-axis is the formation enthalpy in eV/atom relative to the ground states of the corresponding elemental allotropes.}}
\end{figure*}

\begin{figure*}
\epsfig{file=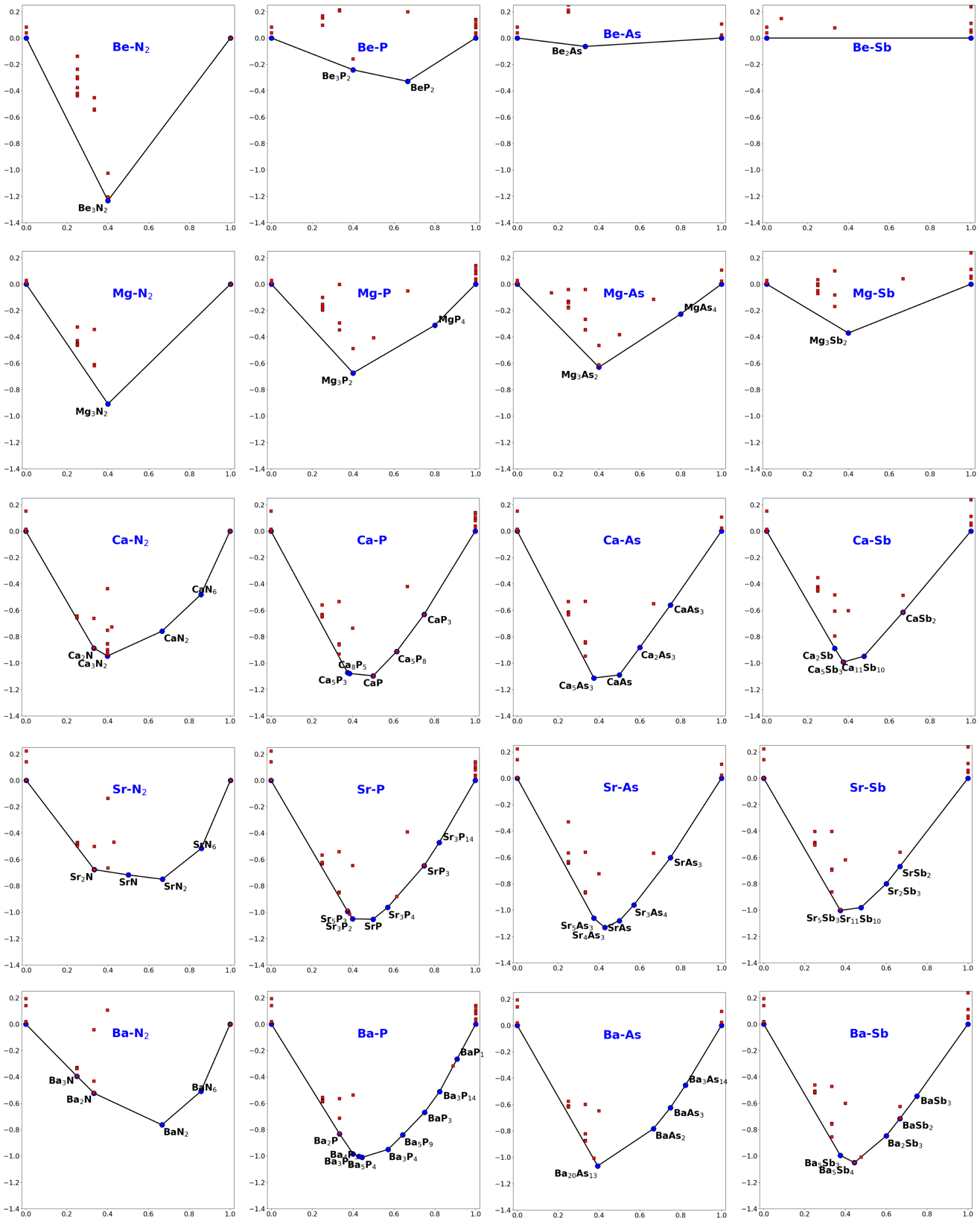, width=0.9\textwidth}
\caption{\label{ht3} {The computed convex hull diagrams of group IIA-VA binary systems at PBE level based with the same parameters setting used in Materials Project. In each diagram, the x-axis is composition ratio, and the y-axis is the formation enthalpy in eV/atom relative to the ground states of the corresponding elemental allotropes.}}
\end{figure*}

We investigated the thermodynamic stability of A3B compounds in other binary systems obtained by chemical substitutions from in group IIB-IVA (Fig. \ref{ht1}), IA-VIA (Fig. \ref{ht2}) and IIA-VA (Fig. \ref{ht3}). We selected the best ten structures by fitness ranking from CSP runs on both Rb-O and Cs-O, and then applied chemical substitution to all these binaries. We also included the prototypical structures of several well-known electrides, such as A$_2$B, A$_5$B$_3$, A$_8$B$_5$ in our study. For each structure, we performed the geometry optimization and energy calculation by using the same parameters sets (MPRelaxSet and MPStaticSet) as used in Materials Project (https://materialsproject.org/docs/calculations). Combining with the available data in the Materials Project3, we computed the final convex hull diagram for each binary system under investigation at PBE level. This analysis identified several other binary systems including alkaline metal oxides and alkaline earth metal nitrides in which the A$_3$B compounds are thermodynamically stable. Therefore, we performed more accurate calculations at optPBE level on these two sets of binaries. The results are summarized in Table \ref{tablea3b}.

\begin{table}
\caption{Stability of monoclinic($C$2/$m$) and hexagonal($P6_3$/$mcm$) A$_3$B compounds at both PBE and optPBE levels. Here the $\times$ symbol denotes the structure is unstable, while the $\surd$ symbol means the structure is stable.}
\begin{tabular}{clcc}
\hline
\hline
\multicolumn{1}{l}{Compound} & Structure    & PBE     & Opt-PBE  \\ \hline
\multirow{2}{*}{Li$_3$O}     & $C$2/$m$     & $\times$& $\times$ \\ 
                             & $P6_3$/$mcm$ & $\times$& $\times$ \\ 
\multirow{2}{*}{Na$_3$O}     & $C$2/$m$     & $\times$& $\times$ \\ 
                             & $P6_3$/$mcm$ & $\times$& $\times$ \\ 
\multirow{2}{*}{K$_3$O}      & $C$2/$m$     & $\times$& $\surd$  \\ 
                             & $P6_3$/$mcm$ & $\times$& $\times$ \\ 
\multirow{2}{*}{Rb$_3$O}     & $C$2/$m$     & $\surd$ & $\surd$  \\ 
                             & $P6_3$/$mcm$ & $\times$& $\times$ \\ 
\multirow{2}{*}{Cs$_3$O}     & $C$2/$m$     & $\times$& $\times$ \\ 
                             & $P6_3$/$mcm$ & $\surd$ & $\surd$  \\
\multirow{2}{*}{Mg$_3$N}     & $C$2/$m$     &$\times$& $\times$\\ 
                             & $P6_3$/$mcm$ &$\times$& $\times$\\ 
\multirow{2}{*}{Ca$_3$N}     & $C$2/$m$     &$\times$& $\times$\\ 
                             & $P6_3$/$mcm$ &$\times$& $\times$\\ 
\multirow{2}{*}{Sr$_3$N}     & $C$2/$m$     &$\times$& $\times$\\ 
                             & $P6_3$/$mcm$ &$\times$& $\times$\\ 
\multirow{2}{*}{Ba$_3$N}     & $C$2/$m$     &$\times$& $\times$ \\ 
                             & $P6_3$/$mcm$ &$\times$& $\surd$ \\
\hline
\hline
\end{tabular}
\label{tablea3b}
\end{table}

\end{document}